# Evolutionary Approaches to Creativity

Liane Gabora
University of British Columbia

Scott Barry Kaufman
Yale University

*Note: There may be minor differences between this draft and the final version accepted for publication.*

Address for Correspondence:
L. Gabora <liane.gabora@ubc.ca>
Department of Psychology
University of British Columbia
Okanagan campus, 3333 University Way
Kelowna BC, V1V 1V7
CANADA

# 1. INTRODUCTION

Many species engage in acts that could be called creative (Kaufman & Kaufman, 2004). However, human creativity is unique in that it has completely transformed the planet we live on. We build skyscrapers, play breathtaking cello sonatas, send ourselves into space, and even decode our own DNA. Given that the anatomy of the human brain is not so different from that of the great apes, what enables us to be so creative? Recent collaborations at the frontier of anthropology, archaeology, psychology, and cognitive science are culminating in speculative but increasingly sophisticated efforts to piece together the answer to this question. Examining the skeletons of our ancestors gives cues as to anatomical constraints that hindered or made possible various kinds of creative expression. Relics of the past have much to tell us about the thoughts, beliefs, and creative abilities of the people who invented and used them. How the spectacular creativity of humans came about is the first topic addressed in this chapter.

Studies at the intersection of creativity and evolution are not limited to investigations into the biological evolution of a highly creative species. Creative ideas themselves might be said to evolve through culture. Human creativity is distinctive because of the adaptive and open-ended manner in which change accumulates. Inventions build on previous ones in ways that enhance their utility or aesthetic appeal, or make them applicable in different situations. There is no *a priori* limit to how a creative idea might unfold over time. A cartoon character can inspire the name and logo for a hockey team (the Mighty Ducks), which might in turn inspire toys, cereal shapes, cigarette lighter designs, or for that matter work its way into an academic book chapter. It is this proclivity to take an idea and make it our own, or 'put our own spin on it', that makes creative ideas appear to evolve. The next section of this chapter investigates in what sense creative ideas evolve through culture.

Finally, we address the question of *why* creativity evolved. What forces supported the evolution of creativity? Does being creative help us live longer, or attract mates? Perhaps creative projects can sometimes *interfere* with survival and reproductive fitness; are there non-biological factors that compel us to create? This is a third topic addressed in this chapter.

# 2. THE BIRTH OF HUMAN CREATIVITY

Looking at an artifact that was fashioned thousands or millions of years ago is an awe-inspiring experience because it gives us a glimpse into the lives and worldviews of our earliest ancestors. To be sure, creative works disintegrate. The farther back in time we look for signs of creativity, the fewer creative works of that time remain with us today. But by corroborating theory and data from different fields, we are on our way toward putting together a coherent picture of how and when the creative abilities of humans arose.

We begin this section by examining the archaeological evidence for the earliest indications of human creativity, and anthropological evidence for concurrent changes in the size and shape of the cranial cavity. We then examine various hypotheses that have been put forward to explain these data.

### *2.1. The Earliest Evidence of Human Creativity: Homo habilis*

It is generally agreed that ancestral humans started diverging from ancestral apes approximately six million years ago. The first Homo lineage, *Homo habilis*, appeared approximately 2.4 million years ago in the Lower Palaeolithic. The earliest known human inventions, referred to as *Oldowan* artifacts (after Olduvai Gorge, Tanzania, where they were first found), are widely attributed to *Homo habilis* (Semaw *et al*., 1997), although it is possible that they were also used by late australopithecenes (de Baune, 2004). They were simple, mostly single faced stone tools, pointed at one end (Leakey, 1971). These tools were most likely used to split fruits and nuts (de Baune, 2004), though some of the more recent ones have sharp edges, and are found with cut-marked bones, suggesting that they were used to sharpen wood implements and butcher small game (Leakey, 1971; Bunn & Kroll, 1986).

These early tools were functional, but simple and unspecialized; by our standards they were not very creative. Feist (2007) refers to the minds of these early hominds as *pre-representational*, suggesting that hominids at this time were not capable of forming representations that deviated from their concrete sensory perceptions; their experience was tied to the present moment. Similarly, Mithen (1996) refers to minds at this time as possessing *generalized intelligence*, reflecting his belief that domain general learning mechanisms, such as Pavlovian conditioning and implicit learning (e.g., Reber, 1993), predominated.

Nevertheless, the early tools of this period mark a momentous breakthrough for our species. Today we are accustomed to seeing everywhere the outcomes of what began as a spark of insight in someone's mind, but when the world consisted solely of naturally-formed objects, the capacity to imagine something and turn it into a reality may well have seemed almost magical. As de Baune (2004) puts it, "the moment when a hominin ...produced a cutting tool by using a thrusting percussion ...marks a break between our predecessors and the specifically human" (p. 142).

### 2.2. *The Adaptive Larger-Brained Homo erectus*

*Homo habilis* persisted from approximately 2.4 to 1.5 million years ago. Approximately 1.8 million years ago, Homo erectus appeared, followed by *Homo ergaster, archaic Homo sapiens*, and *Homo neanderthalensis*. The size of the *Homo erectus* brain was approximately 1,000 cc, about 25% larger than that of *Homo habilis*, and 75% the cranial capacity of modern humans (Aiello, 1996; Ruff *et al*., 1997; Lewin, 1999). *Homo erectus* exhibit many indications of enhanced ability to creatively adapt to the environment to meet the demands of survival, including sophisticated, task-specific stone hand axes, complex stable seasonal habitats, and long-distance hunting strategies involving large game. By 1.6 Ma, *Homo erectus* had dispersed as far as Southeast Asia, indicating the ability to adjust lifestyle to vastly different climates (Anton & Swisher, 2004; Cachel & Harris, 1995; Swisher, Curtis, Jacob, Getty, & Widiasmoro, 1994; Walker & Leakey, 1993). In Africa, West Asia, and Europe, by 1.4 Ma *Homo erectus* developed the Aschulean handaxe (Asfaw et al., 1992), a do-it-all tool that may even have had some function as a social status symbol (Kohn & Mithen, 1999). These symmetrical biface tools probably required several stages of production, bifacial knapping, and considerable skill and spatial ability to achieve their final form.

Though the anatomical capacity for language was present by this time (Wynn, 1998), verbal communication is thought to have been limited to (at best) pre-syntactical proto-language (Dunbar, 1996). Additionally, while humans may have for the first time been capable of representing an idea once the object was no longer being present, such representations were more likely to be visual rather than verbal (Feist, 2006).

Also, thought during this time period was most likely only first-order; the capacity for thinking about thinking (i.e., metacognition) had not yet developed. Some suggest that this period witnessed the emergence of domains of knowledge associated with social, physical, biological, and quantitative concepts (Feist, 2006), although Mithen (1996) argues that such knowledge was at this time probably encapsulated.

### *2.3. Possible Explanations for the Onset of Human Creativity*

It has been suggested that these early archaeological finds do not reflect any underlying biological change, but were simply a response to climactic change (Richerson & Boyd, 2000). However given the above-mentioned significant increase in cranial capacity, it seems parsimonious to posit that this dramatic encephalization allowed a more sophisticated mode of cognitive functioning, and is thus at least partly responsible for the appearance of cultural artifacts (and the beginnings of an archaeological record).

There are multiple versions of the hypothesis that the onset of the archaeological record reflects an underlying cognitive transition. One suggestion is that the appearance of archaeological novelty is due to the onset of the capacity to imitate (Dugatkin, 2001), or onset of *theory of mind*—the capacity to reason about mental states of others (Premack & Woodruff, 1978). Although these hypotheses may explain how new ideas, once in place, spread from one individual to another, they are inadequate as an explanation of the enhanced capacity for coming up with new ideas in the first place. Moreover, other species possess theory of mind (Heyes, 1998), and imitate, (Byrne & Russon, 1998; Darwin, 1871) yet do not compare to hominids with respect to creativity.

Yet another proposal is that *Homo* underwent a transition at this time from an *episodic mode* of cognitive functioning to a *mimetic mode* (Donald, 1991). The episodic mind of *Homo habilis* was sensitive to the significance of episodes, and could encode them in memory and coordinate appropriate responses, but could not voluntarily access them independent of cues. The enlarged cranial capacity of *Homo erectus* enabled it to acquire a mimetic form of cognition, characterized by possession of what Donald (1991) refers to as a 'self-triggered recall and rehearsal loop', or SRRL. The SRRL enabled hominids to voluntarily access memories independent of cues, and thereby act out events that occurred in the past, or that could occur in the future (indeed the term mimetic is derived from the word 'mime'). Thus not only could the mimetic mind temporarily escape the here and now, but through gesture it could bring about a similar escape in other minds. The SRRL also enabled hominids to engage in a stream of thought, such that attention is directed away from the external world toward ones' internal model of it, and one thought or idea evokes another, revised version of it, which evokes yet another, and so forth recursively. Finally, the SRRL enabled the capacity to evaluate and improve motor acts through repetition or rehearsal, and adapt them to new situations, resulting in more refined artifacts and survival tactics.

It seems reasonable that a larger brain might be more likely to engage in self-triggered recall and rehearsal, but Donald's scenario becomes even more plausible when considered in light of the structure and dynamics of associative memory (Gabora, 1998, 2003, 2007). We know that neurons are sensitive to *subsymbolic microfeatures*—primitive attributes of a stimulus, such as a sound of a particular pitch or a line of a particular orientation. Episodes etched in memory are *distributed* across a bundle or cell assembly of these neurons, and likewise, each neuron participates in the encoding of many episodes. Finally, memory is *content-addressable*, such that similar stimuli activate and get encoded in overlapping distributions of neurons. With a larger brain, episodes are encoded in more detail, i.e. there is a transition from a more coarse-grained to a more fine-grained memory. This means that there are more ways in which distributions can overlap, and thus more routes by which one can evoke another, thus providing an anatomical basis for self-triggered recall and rehearsal, and the forging of creative connections. It also paved the way for a more integrated internal model of the world, or worldview.

## 3. OVER A MILLION YEARS OF CREATIVE STASIS

The handaxe persisted as the almost exclusive tool of choice for over a million years, spreading by 500,000 years ago into Europe, where was it used until about 200,000 years ago. Indeed during this period not only is there almost no change in tool design, but little evidence of creative insight of any kind, with the exception of the first solid evidence for controlled use of fire, by 800,000 years ago in the Levant (Goren-Inbar, et al., 2004).

### *3.1. A Second Increase in Brain Size*

Between 600,000 and 150,000 years ago there was a second spurt in brain enlargement (Aiello 1996; Ruff *et al*. 1997). But although *anatomically* modern humans had arrived, *behavioral* modernity had not. It would make our story simple if the increase in brain size coincided with the burst of creativity in the Middle/Upper Paleolithic (Bickerton, 1990; Mithen, 1998), to be discussed shortly. It does correspond with the revolutionary advancement of the Levallois flake, which came into prominence approximately 250,000 years ago in the Neanderthal line. But although one sees in the artifacts of this time the germ of modern day representational thought, it is clear that cognitive processes are still primarily first-order—tied to concrete sensory experience—rather than second-order—derivative, or abstract. Leakey (1984) writes of anatomically modern human populations in the Middle East with little in the way of culture, and concludes "The link between anatomy and behavior therefore seems to break" (p. 95).

It may be that this second spurt in brain size exerted an impact on expressions of creativity that leave little trace in the archaeological record such as ways of coping with increasing social complexity, for example, manipulating competitors for purposes of survival and reproduction (Baron-Cohen, 1995; Byre & Whiten, 1988; Humphrey, 1976; Whiten, 1991; Whiten & Byrne, 1997; Wilson et al., 1996, Dunbar, 1996). Another possible reason for the rift between anatomical and behavioral modernity is that while genetic mutations necessary for cognitive modernity arose at this time, the fine-tuning of the nervous system to capitalize on these genetic changes took longer, or required certain environmental conditions to be manifested. It is worth noting that other periods of revolutionary innovation, such as the Holocene transition to agriculture and the modern Industrial Revolution, occurred long after the biological changes that made them cognitively possible. Yet another possibility (to be elaborated shortly) is that the explosion of creativity in the Middle/Upper Paleolithic occurred earlier, or more gradually, than originally believed.

## 4. THE SPECTACULARLY CREATIVE MIND OF MODERN HUMANS

The European archaeological record indicates that a truly unparalleled cultural transition occurred between 60,000 and 30,000 years ago at the onset of the Upper Paleolithic (Bar-Yosef, 1994; Klein, 1989a; Mellars, 1973, 1989a, 1989b; Soffer, 1994; Stringer & Gamble, 1993). Considering it "evidence of the modern human mind at work," Richard Leakey (1984:93-94) describes the Upper Palaeolithic as follows: "unlike previous eras, when stasis dominated, ... [with] change being measured in millennia rather than hundreds of millennia." Similarly, Mithen (1996) refers to the Upper Paleaolithic as the 'big bang' of human culture, exhibiting more innovation than in the previous six million years of human evolution. At this time that we see the more or less simultaneous appearance of traits considered diagnostic of behavioral modernity. It marks the beginning of a more organized, strategic, season-specific style of hunting involving specific animals at specific sites, elaborate burial sites indicative of ritual and religion, evidence of dance, magic, and totemism, the colonization of Australia, and replacement of Levallois tool technology by blade cores in the Near East. In Europe, complex hearths and many forms of art appeared, including naturalistic cave paintings of animals, decorated tools and pottery, bone and antler tools with engraved designs, ivory statues of animals and sea shells, and personal decoration such as beads, pendants, and perforated animal teeth, many of which may have indicated social status (White, 1989a, 1989b). White (1982:176) also writes of a "total restructuring of social relations." What is perhaps most impressive about this period is not the novelty of any particular artifact but that the overall pattern of cultural change is cumulative; more recent artifacts resemble older ones but have modifications that enhance their appearance or functionality. This is referred to as the *ratchet effect* (Tomasello, 1993), and it appears to be uniquely human (Donald, 1998).

Whether this period was a genuine revolution culminating in behavioral modernity is hotly debated because claims to this effect are based on the European Palaeolithic record, and largely exclude the African record (McBrearty & Brooks, 2000); Henshilwood & Marean, 2003). Indeed, most of the artifacts associated with a rapid transition to behavioral modernity at 40–50,000 years ago in Europe are found in the African Middle Stone Age tens of thousands of years earlier. These include blades and microliths, bone tools, specialized hunting, long distance trade, art and decoration (McBrearty & Brooks, 2000), the Berekhat Ram figurine from Israel (d'Errico & Nowell, 2000), and an anthropomorphic figurine of quartzite from the Middle Ascheulian (ca. 400 ka) site of Tan-tan in Morocco (Bednark, 2003). Moreover, gradualist models of the evolution of cognitive modernity well before the Upper Palaeolithic find some support in archaeological data (Bahn, 1991; Harrold, 1992; Henshilwood & Marean, 2003; White, 1993; White *et al*., 2003). If modern human behaviors were indeed gradually assembled as early as 250–300,000 years ago, as McBrearty and Brooks (2000) argue, it pushes the transition into alignment with the most recent spurt in human brain enlargement. However the traditional and currently dominant view is that modern behavior appeared in Africa between 50,000 and 40,000 years ago due to biologically evolved cognitive advantages, and spread replacing existing species, including the Neanderthals in Europe (e.g., Ambrose, 1998; Gamble, 1994; Klein, 2003; Stringer & Gamble, 1993). Thus from this point onward there was only one hominid species: the modern *Homo sapien*.

Despite lack of overall increase in cranial capacity, the prefrontal cortex, and particularly the orbitofrontal region, increased significantly in size (Deacon, 1997; Dunbar, 1993; Jerison, 1973; Krasnegor, Lyon, and Goldman-Rakic, 1997; Rumbaugh, 1997) and it was likely a time of major neural reorganization (Klein, 1999; Henshilwood, d'Errico, Vanhaeren, van Niekerk, and Jacobs, 2000; Pinker, 2002). These brain changes most likely gave rise to what Feist (2006) refers to as "meta-representational thought" or the ability to reflect on representations and think about thinking. Along similar lines, Dennet (1976) argues that an important transition in the evolution of *Homo sapiens* is from first-order intentionality to second-order intentionality. According to Dennet, a first-order intentional system has beliefs and desires, but cannot reflect on those beliefs and desires. A second-order intentional system by contrast has beliefs and desires about the beliefs and desires of themselves and others.

At any rate, it is accepted that the Middle/Upper Palaeolithic was a period of unprecedented creativity. How and why did it occur? What kind of cognitive processes were involved?

### *4.1. Cognitive Explanations*

Let us now review the most popular hypotheses for what kind of biologically evolved cognitive advantages gave rise to behavioral modernity at this time.

#### *4.1.1 Advent of Syntactic Language*

It has been suggested that humans underwent at this time a transition from a predominantly gestural to a vocal form of communication (Corballis, 2002). Although due the ambiguity of the archaeological evidence we may never know exactly when language began (Bednarik, 1992:30; Davidson & Noble, 1989), most scholars agree that while earlier Homo and even Neanderthals may have been capable of primitive proto-language, the grammatical and syntactic aspects emerged at the start of the Upper Palaeolithic (Aiello & Dunbar, 1993; Bickerton, 1990, 1996; Dunbar, 1993, 1996; Tomasello, 1999). Carstairs-McCarthy (1999) presents a modified version of this proposal, suggesting that although some form of syntax was present in the earliest languages, most of the later elaboration, including recursive embedding of syntactic structure, emerged in the Upper Paleolithic. Syntax enabled language to become general-purpose, and put to use in a variety of situations. It enhanced not just the ability to communicate with others, spread ideas from one individual to the next, and collaborate on creative projects (thereby speeding up cultural innovation), but also the ability to think things through precisely for oneself and manipulate ideas in a controlled, deliberate fashion (Reboul, 2007).

#### *4.1.2 Cognitive Fluidity*

Fauconnier & Turner (2002) propose that the exceptional creativity of the Middle/Upper Paleolithic was due to onset of the capacity to blend concepts, which facilitated the weaving of experiences into stories and parables. A similar explanation is put forward by Mithen (1996), drawing on the evolutionary psychologist's notion of massive modularity (Buss, 1999, 2004; Buss *et al*., 1994; Cosmides & Tooby, 2002;

Dunbar et al., 1994; Rozin, 1976; for an extensive critique see Buller, 2005). Mithen suggests that the creativity of the modern mind arose through the onset of *cognitive fluidity*, resulting in the *connecting* of what were previously encapsulated (functionally isolated) brain modules devoted to natural history, technology, social processes, and language. This he claims gave us the ability to map, explore, and transform conceptual spaces, referring to Boden's (1990) definition of a conceptual space as a 'style of thinking—in music, sculpture, choreography, chemistry, *etc*.' Sperber (1994) proposes that the connecting of modules involved a special module, the 'module of meta-representation' or MMR, which contains 'concepts of concepts', and enabled cross-domain thinking, and particularly analogies and metaphors.

Note that the notion of modules amount to an explicit high-level compartmentalization of the brain for different tasks. However this kind of division of labor—and the ensuing creativity—would emerge unavoidably as the brain got larger *without* explicit high-level compartmentalization, due to the sparse, distributed, content-addressable manner in which neurons encode information (Gabora, 2003). Neurons are tuned to respond to different subsymbolic microfeatures, and there is a systematic relationship between the content of a stimulus and the distributed set of neurons that respond to it, such that neurons that respond to similar microfeatures are near one another (Churchland & Sejnowski, 1992; Smolensky, 1988). Thus, as the brain got larger and the number of neurons increased, and the brain accordingly responded to a greater variety of subsymbolic microfeatures, neighboring neurons tended to respond to microfeatures that were more similar, and distant neurons tended to respond to microfeatures that were more different. There were more ways in which distributed representations could overlap and creative connections be made. Thus a weak modularity of sorts emerges naturally at the level of the neuron without any explicit high-level compartmentalization going on, and it need not necessarily correspond to how humans carve up the world, i.e. to categories such as natural history, technology, and so forth. Moreover, explicit connecting of modules is not necessary for creative connections to be made; all that is necessary is that the relevant domains or modules be simultaneously accessable (Gabora, 2003).

#### *4.1.3 Symbolic Reasoning*

Another suggestion is that the creativity of the Middle/Upper Paleaolithic was due to the emergence of an ability to internally represent complex, abstract, internally coherent systems of meaning, including symbols and the *causal relationships* amongst them (Deacon, 1997). Deacon believes this colored our existence by making us view objects and people in terms of the roles they could play in stories, and the point or meaning they could potentially have, or participate in.

Indeed, onset of the capacity for symbolic representation certainly plays a fundamental role in the mental life of modern humans. On the other hand though, those versed in the creativity literature tend to think that the intuitive, divergent, associative processes with which we unearth *relationships of correlation* play at least as great a role.

#### *4.1.4 Contextual focus*

The above proposals for what kind of cognitive change could have led to the Upper Paleolithic transition stress different aspects of cognitive modernity. Acknowledging a possible seed of truth in each of them we begin to converge toward a common (if more complex) view. Conceptual blending is characteristic of *divergent thought*, which tends to be automatic, associative, intuitive, and diffuse. This is quite different from the *convergent thought* stressed by Deacon, which tends to be logical, controlled, effortful, and reflective. Converging evidence suggests that the modern mind engages in both (Arieti, 1976; Ashby & Ell, 2002; Freud, 1949; Guilford, 1950; James, 1890/1950; Johnson-Laird, 1983; Kris, 1952; Neisser, 1963; Piaget, 1926; Rips, 2001; Sloman, 1996; Stanovich & West, 2000; Werner, 1948; Wundt, 1896). This is sometimes referred to as the dual-process theory of human cognition (Chaiken & Trope, 1999; Evans & Frankish, in press) and it is consistent with current theories of creative cognition (Finke, Ward, & Smith, 1992; Gabora, 2000, 2002, 2003, under revision; Smith, Ward, & Finke, 1995; Ward, Smith, & Finke, 1999). Divergent processes are hypothesized to occur during idea generation, while convergent processes predominate during the refinement, implementation, and testing of an idea. Hence it has been proposed that the Paleolithic transition reflects a mutation to the genes involved in the fine-tuning of the biochemical mechanisms underlying the capacity to subconsciously shift between these modes, depending on the situation, by varying the specificity of the activated cognitive receptive field (Gabora, 2003, 2007; for similar ideas see Howard-Jones & Murray, 2003; Martindale, 1995). This is referred to as *contextual focus*[1] because it requires the ability to focus or defocus attention in response to the context or situation one is in. Defocused attention, by diffusely activating a broad region of memory, is conducive to divergent thought; it enables obscure (but potentially relevant) aspects of the situation thus come into play. Focused attention is conducive to convergent thought; memory activation is constrained enough to hone in and perform logical mental operations on the most clearly relevant aspects. Thus in an analytic mode of thought the concept GIANT might only activate the notion of large size, whereas in an associative mode the giants of fairytales might come to mind. Once it was possible to shrink or expand the field of attention, and thereby tailor one's mode of thought to the demands of the current situation, tasks requiring either convergent thought (*e.g.* mathematical derivation), divergent thought (*e.g.* poetry) or both (*e.g.* technological invention) could be carried out more effectively. When the individual is fixated or stuck, and progress is not forthcoming, defocusing attention enables the individual to enter a more divergent mode of thought, and working memory expands to include peripherally related elements of the situation. This continues until a potential solution is glimpsed, at which point attention becomes more focused and thought becomes more convergent, as befits the fine-tuning and manifestation of the creative work.

Thus the onset of contextual focus would have enabled the hominid to adapt ideas to new contexts or combine them in new ways through divergent thought, and fine-tune these strange new combinations through convergent thought. In this way the fruits of one mode of thought provide the ingredients for the other, culminating in a more fine-grained internal model of the world.

#### *4.1.1. Shifting Between Implicit and Explicit Thought*

In a similar vein, Feist (2007) suggests that cognitive fluidity enabled hominids to move not just horizontally between domains (as Mithen (1996) suggests), but also vertically between implicit and explicit modes of thought, allowing for the ability to make the broad associations seen in high levels of creativity, and to arrive at novel and useful solutions and ideas.

---

[1] In neural net terms, contextual focus amounts to the capacity to spontaneously and subconsciously vary the shape of the activation function, flat for divergent thought and spiky for analytical.

Indeed, while explicit cognition is often equated with our highest cognitive abilities, the set of autonomous subsystems that underlie implicit cognition (e.g., Stanovich, 2004) also plays an important role in creative cognition. One particularly important type of implicit processing is implicit learning, which encompasses the ability to automatically and nonconsciously detect complex regularities, contingencies, and covariances in our environment. Implicit learning is a fundamental aspect of our humanness, and shared by our most distant ancestors. Even among modern humans it plays a significant role in structuring our skills, perceptions, and behavior (Berry & Broadbent, 1988; Cleeremans & Jiménez, 1997; Hassin, Uleman, & Bargh, 2004; Lewicki & Hill, 1987; Lewicki, Czyzewska, & Hoffman, 1987; Kaufman, 2007; McGeorge & Burton, 1990; A. Reber, 1967; 1993; P. Reber & Kotovsky, 1997; Squire & Frambach, 1990). Indeed it is probably implicit not explicit cognition that is responsible for creative insight (Bowers et al, 1995; Kaufman, 2008).

A contributing factor to the emergence of the ability to shift modes of thought may have been the expansion of the prefrontal cortex, and the associated executive functions and enhanced working memory[2] capacity that came with the expansion. Enhanced working memory allowed humans more control over their focus of attention so as to maintain task goals in the presence of interference. Indeed, individual differences in working memory capacity are strongly related to fluid intelligence among modern humans (Conway, Jarrold, Kane, & Miyake, 2007; Engle, Tuholski, Laughlin, & Conway, 1999; Kane, Hambrick, & Conway, 2005; Kaufman, DeYoung, Gray, Brown, & Mackintosh, under revision).

Therefore, the fruits of implicit processes come to conscious awareness only once they have been honed into a form in which we can mentally operate on them. Then executive functions, associated with the growth of the prefrontal cortex, and aided by the use of syntactical language, can reflectively and explicitly manipulate the once implicit associations to achieve goals that may sometimes diverge from the rigid gene level goals of our distant ancestors (Stanovich, 2005).

### *4.2. A Return to the Lag between Anatomical and Behavioral Modernity*

Let us return briefly to the question of why the burst of creativity in the Upper Paleolithic occurred well after the second rapid increase in brain size approximately 500,000 years ago. A larger brain provided more room for episodes to be encoded, and particularly more association cortex for connections between episodes to be made, but it doesn't follow that this increased brain mass could straightaway be optimally navigated. There is no reason to expect that information from different domains (whether strongly modular or weakly modular) would immediately be compatible enough to coexist in a stream of thought, as in the production of a metaphor. It is reasonable that it took time for the anatomically modern brain to fine-tune how its components 'talk' to each other such that different items could be blended together or recursively revised and recoded in a coordinated manner (Gabora, 2003). Only then could the full potential of the large brain be realized. Thus the bottleneck may not have been sufficient brain size but sufficient sophistication in the *use of* the memory already available, through contextual focus, or shifting between implicit and explicit thought.

[2] Working memory is the ability to maintain, update, and manipulate information in an active state.

### *4.3. The Multi-Layered Mind and 'Recent' Creative Breakthroughs*

Several researchers emphasize that the modern human mind consists of various 'kinds of minds' layered on top of one another (Reber, 1989, 1993; Reber & Allen, 2000; Dennett, 1995, 1996). According to these accounts, these multiple minds are continuously operative, giving rise to many internal and external conflicts amongst members of our species, as well as contributing to our most distinctly human intellectual and creative accomplishments.

According to Arthur Reber, implicit cognition is evolutionarily older than explicit cognition. Reber speculates that the arrival of explicit cognition, which includes processes of hypothesis-guided learning and deduction, have not modified the older mechanisms of implicit learning that continue to function independently. Dennett (1996) elaborates on this, arguing that the various kinds of minds that exist in modern Homo sapiens differ to the extent to which each is rigidly tied to gene level goals (as opposed to goals held by the individual, or 'vehicle'). For instance, our earliest evolved minds consist of behavioral patterns that are prewired and act like reflexes, and are more tied to gene level goals, while our later evolving capacities for reflection and deliberate reasoning gave humans the flexibility to override gene level goals in the service of vehicle level goals (Stanovich, 2005).

Of course the story of how human creativity evolved does not end with the arrival of anatomical and behavioral modernity. The end of the ice age around 10-12,000 years ago witnessed the beginnings of agriculture and the invention of the wheel. Written languages developed around 5-6,000 years ago, and approximately 4,000 years ago astronomy and mathematics appear on the scene. We see the expression of philosophical ideas around 2,500 years ago, invention of the printing press 1,000 years ago, and the modern scientific method about 500 years ago. And the past 100 years have yielded a technological explosion that has completely altered the daily routines of humans (as well as other species), the consequences of which remain to be seen.

## 5. CREATIVITY AND CULTURAL EVOLUTION

We have examined how the *capacity* for creativity evolved over millions of years. In this section we explore the possibility that creative ideas *themselves* evolve through culture, in the sense that they exhibit 'descent with modification', or incremental adaptation to the constraints of their environment. (A related idea is that the creative process not at the cultural level but within the mind of one individual is Darwinian; this is discussed in the chapter on theories, this volume.)

### *5.1. Creative Cultural Change as a Darwinian Process*

It has been proposed that the process by which creative ideas change over time as they pass from person to person can be described in Darwinian terms (Aunger, 2000; Blackmore, 1999; Boyd & Richerson, 1985; Cavalli-Sforza & Feldman, 1981; Dawkins, 1975; Durham, 1991). This approach is sometimes referred to as 'dual inheritance theory', the idea being that we inherit cultural as well as biological information, and the units of cultural information are sometimes referred to as 'memes'. The rationale is clear; since natural selection is useful for explaining the astonishing creativity of nature, perhaps it is also useful for explaining the astonishing creativity of human culture. There are many parallels between the two. Clearly new inventions build on existing ones, but it isn't just the *cumulative* nature of human creativity that is reminiscent of biological evolution. Cumulative change is after all rather easy to come by; in the days of taping music, each time a tape was copied it became cumulatively more scratched. The creativity of human cultures is reminiscent of biological evolution because of the *adaptive and open-ended manner* in which change accumulates. New inventions don't just build on old ones, they do so in ways that meet our needs and appeal to our tastes, and as in biological evolution there is no limit to how any particular invention or creative work may inspire or influence other creative works. Moreover, culture generates phenomena observed in biological evolution, such as drift[3] and niches[4] (Bentley et al., 2004; Gabora, 1995, 1997). A theory that encompasses the two would put us on the road to uniting the social sciences with the biological sciences.

In order to see how Darwinian theory might be applied to the evolution of creativity ideas in culture, let us examine what kind of process natural selection can describe, and how it works. The paradox faced by Darwin and his contemporaries was the following: how does biological change accumulate when traits acquired over an organism's lifetime are obliterated? For example, a rat whose tail is cut off does not give birth to rats with cut-off tails; the rat lineage loses this trait. Note that this kind of continual 'backtracking' to an earlier state is unique to biology; if for example, an asteroid crashes into a planet, the planet cannot revert to the state of having not had the asteroid crash into it.[5]

Darwin's genius was to explain how living things adapt over time despite that new modifications keep getting discarded, by looking from the level of the individual to the level of the *population* of interbreeding individuals. He realized that individuals who are better equipped to survive in their given environment leave more offspring (are 'selected'). Thus, although their *acquired* traits are discarded, their *inherited* traits (loosely speaking, the traits they were born with) are more abundant in the next generation. Over generations this can lead to substantial change in the distribution of traits across the population as a whole. Natural selection was not put forth to explain how biological novelty originates. It assumes random variation of heritable traits, and provides an explanation for population-level change in the *distribution* of variants.

---

[3] Drift refers to changes in the relative frequencies of variants through random sampling from a finite population. It is the reason why variation is reduced in reproductively isolated populations such as those living on a small island. Drift has been shown to occur in a culture context with respect to such things as baby names and dog breed preferences (Neiman 1995; Madsen et al. 1999; Bentley et al. 2004). In a computer model of cultural evolution, the smaller the society of artificial agents, the lower the cultural diversity (Gabora, 1995).

[4] Just as the biological evolution of rabbits created niches for species that eat them and parasitize their guts, the cultural evolution of cars created niches for seat belts and gas stations (Gabora, 1997, 1998).

[5] Although Darwin observed that this was the case, he did not know why. We now know that the reason acquired traits are not inherited in biology is that organisms replicate using a template—a self-assembly code that is both actively transcribed to produce a new individual, and passively copied to ensure that the new individual can itself reproduce.

We now ask: can natural selection similarly explain the process by which creative ideas evolve through culture? A first thing that can be noted is that the problem for which natural selection was put forward as a solution does not exist with respect to culture (Gabora, 2008). That is, there is no sense in which the components of creative ideas cyclically accumulate and then get discarded at the interface between one generation and the next. For example, unlike the chopped off tail which does not get transmitted to offspring, once someone invented the spout on a teapot, teapots could forever after have spouts. One might ask if Darwin's solution is nevertheless applicable; might processes outside of biology evolve through selection even if selection was originally advanced as a solution to a paradox that is unique to biology? The problem is that since acquired change can accumulate orders of magnitude faster than inherited change, if it is *not* getting regularly discarded, it quickly swamps the population-level mechanism of change identified by Darwin. This is particularly the case with respect to creative ideas since they do not originate through random processes—or even processes prone to canceling one another out—but through strategic or implicit, intuitive processes, making use of the associative structure of memory.

Darwinian approaches to culture posit that the basic units of this second Darwinian process are discrete elements of culture that pass from one person to another intact except for random change akin to mutation that arises through copying error or biased transmission (preferential copying of high status individuals). Copying error and biased transmission are sources of change that take place at the time an idea spreads from one individual to another, which creativity researchers tend to view as a relatively minor source of creative change compared with cognitive processes such as imagining, planning, analogizing, concept combination, and so forth. The reason that Darwinian theories of culture focus on sources of change that occur when an idea spreads from one individual to another is not accidental; it stems from the fact that natural selection is only of explanatory value to the extent that there is negligible inheritance or transmission of acquired characteristics. This is the case in biology, as we saw with the cut-off tail example; change acquired during an individual's lifetime is not generally passed on to its offspring. As another example, you didn't inherit your mother's tattoo—something she acquired between the time she was born and the time she transmitted genetic material to the next generation.

However, few scholars accept that there is negligible transmission of acquired characteristics in culture. The cultural equivalent of the individual is the creative idea. A new 'generation' begins when this idea is transmitted from person A to person B, and lasts until the idea is transmitted from person B to person C. Any changes to an idea between the time B learned it and the time B expressed it are 'acquired characteristics'. If B mulls the idea over or puts it into her own terms or adapts it to her own framework, the process by which this idea changes cannot be explained by natural selection, because as mentioned earlier, this kind of intra-generational change quickly drowns out the slower inter-generational mechanism of change identified by Darwin; it 'swamps the phylogenetic signal'. The Darwinian perspective on culture therefore leads to a view of the human condition as 'meme hosts', passive imitators and transmitters of prepackaged units of culture, which evolve as separate lineages. To the extent that these lineages 'contaminate one another'; that is, to the extent that we actively and creatively transform elements of culture in ways that reflect our own internal models of the world, altering or combining them to suit our needs, perspectives, or aesthetic sensibilities, natural selection cannot explain cultural change. It has been argued that due to this 'lack of inheritance of acquired characteristics' problem not just the evolution of creative ideas (Gabora, 2005), and the evolution of culture (Gabora, 2004, 2008a), but the evolution of early life itself (Gabora, 2006; Vetsigian et al., 2006), and even of many features of modern life (e.g. Jablonka & Lamb, 2007; Kauffman, 1993; Newman & Müller, 1999; Schwartz, 1999) cannot be described by Darwin's theory of natural selection.

### *5.2. A Non-Darwinian Theory of How Creative Ideas Evolve*

If creative ideas do not evolve through selection, how do they evolve? One possibility is that the evolution of creative ideas through culture is more akin to the evolution of the earliest biological life forms than to present-day DNA-based life (Gabora, 1998, 2000, 2004, 2008). Recent work suggests that early life emerged and replicated through a self-organized process referred to as *autocatalysis*, in which a set of molecules catalyze (speed up) the reactions that generate other molecules in the set, until as a whole they self-replicate (Kauffman, 1993). Such a structure is said to be *autopoietic,* or *self-regenerating*, because the whole is reconstituted through the interactions of the parts (Maturana & Varela, 1980). These earliest precursors of life evolved not through natural selection at the level of the population, like present-day life, but communal exchange of innovation at the individual level (Gabora, 2006; Vetsigian *et al*. 2006). Since replication of these pre-DNA life forms occurred through regeneration of catalytic molecules rather than (as with present day life) by using a genetic self-assembly code, acquired traits were inherited. In other words, their evolution was, like that of culture, Lamarckian.

This had led to the suggestion that it is worldviews that evolve through culture, through the same non-Darwinian process as the earliest forms of life evolved, and creative products such as tools and dances and architectural plans are external manifestations of this process; they reflect the states of the particular worldviews that generate them (Gabora, 1998, 2000, 2004, 2008). The idea is that like these early life forms, worldviews evolve not through natural selection, but through self-organization and communal exchange of innovations. One does not accumulate elements of culture transmitted from others like items on a grocery list, but hones them into a unique tapestry of understanding, a worldview, which like these early life forms is autopoietic in that the whole emerges through interactions amongst the parts. It is *self-mending* in the sense that, just as injury to the body spontaneously evokes physiological changes that bring about healing, events that are problematic or surprising or evoke cognitive dissonance spontaneously evokes streams of thought that attempt to solve the problem or reconcile the dissonance (Gabora, 1999, under revision). Thus Gabora proposes that it is not chance, mutation-like processes that propel creativity, but the self-organizing, self-mending nature of a worldview.

## 6. WHY DID CREATIVITY EVOLVE?

We have discussed how human creativity evolved, and in what sense creative ideas can be said to evolve. We now address a fundamental question: *why* did human creativity evolve?

### *6.1. Creativity as Evolutionary Spandrels*

Some forms of creativity enhance survival and thus reproductive fitness. For example, the invention of weapons most likely evolved as a creative response to a need for protection from enemies and predators. For other forms of creative expression however, such as art and music, the link to survival and reproduction is not so clear-cut. Why do we bother?

Pinker (1997) argues that art, music, humor, fiction, religion, and philosophy are not real adaptations, but evolutionary spandrels: side-effects of abilities that evolved for other purposes. He likens these forms of creativity to cheesecake and pornography—cultural inventions that stimulate our senses in novel ways, but do not improve our biological fitness. Others similarly suggest that the cognitive abilities for planning and remembering important ecological facts extended into our uniquely human capacities for art, poetry, music, story-telling, and humor (Carroll, 1995; Gabora, 2003; Kaufman et al., 2007; McBrearty & Brooks, 2000).

The 'spandrels' explanation assumes that what drives creativity is biological selection forces operating at the individual level, and there is some empirical support for this. Some forms of human creativity, such as art and music, indeed demonstrate the features of a naturally selected adaptation (Dissanayake, 1988, 1992). For one, many forms of creativity are ubiquitous. Although styles differ, every culture creates works of art and music. Second, many forms of creativity are pleasurable for both the artist and the audience, and evolutionarily adaptive behaviors are usually pleasurable. Third, many forms of creativity in humans require effort to produce. Costly behaviors usually do not evolve by accident.

### *6.2. Group Bonding*

Even if creativity is at least in part driven by individual level biological selection forces, other forces may also be at work. Natural selection is believed to operate at multiple levels, including gene-level selection, individual-level survival selection, individual-level sexual selection, kin selection, and group selection. Although there is evidence from archaeology, anthropology, and ethnography that individual-level survival selection plays a key role in human creativity, other levels may have an impact as well. For example, some anthropologists view the function of forms of creativity such as art and music as strengthening a group's social cohesion. For music in particular, Mithen (2006) presents evidence that the melodious vocalizations by our earliest ancestors played an important role in creating and manipulating social relationships through their impact on emotional states.

### *6.3. Sexual Selection*

Miller (2000) argues that group bonding accounts of creativity ignore the possible role of sexual selection in shaping creative behavior, and cannot account for the sexual attractiveness of various forms of creativity. This idea has its roots in Darwin, who once said, "It appears probable that the progenitors of man, either the males or females or both sexes, before acquiring the power of expressing mutual love in articulate language, endeavored to charm each other with musical notes and rhythm (Darwin, 1871, p. 880).

According to the sexual selection account, there is competition to mate with individuals who exhibit creative traits that are (in theory) metabolically expensive, hard to maintain, not easily counterfeited, and highly sensitive to genetic mutation because they are the most reliable indicators of genetic fitness. In recent years, Miller (1998; 2000a; 2000b; 2000c; 2001; Kaufman et al., 2007) has developed and popularized the most elaborated version of this theory. He argues that sexual selection has played a much greater role than natural selection in shaping the most distinctively human aspects of our minds, including storytelling, art, music, sports, dance, humor, kindness, and leadership. He contends that these creative behaviors are the result of complex psychological adaptations whose primary functions were to attract mates, yielding reproductive rather than survival benefits. Miller notes that cultural displays of human creativity satisfy these requirements. According to this account, cultural displays are the result of efforts to broadcast courtship displays to recipients: "art evolved, at least originally, to attract sexual partners by playing upon their senses and displaying one's fitness" (Miller, 2000, p. 267).

Along similar lines, Marek Kohn and Steven Mithen (Kohn, 1999; Kohn & Mithen, 1999) propose what they refer to as the "sexy-hand axe hypothesis". According to this hypothesis, sexual selection pressures may have caused men to produce symmetric hand axes as a reliable indicator of cognitive, behavioral, and physiological fitness. As Mithen (1996) notes, symmetrical hand-axes are often attractive to the modern eye, but require a huge investment in time and energy to make—a burden that makes it hard to explain their evolution in terms of strictly practical, survival purposes. Since hand-axes may be viewed as the first aesthetic artifacts in the archeological record, these products may indeed be the first evidence of sexual selection shaping the emergence of art.

Various scholars have elaborated and clarified this theory. Feist (2001) notes that Miller focuses on sexual selection so much that he excludes the evolution of scientific creativity and technology, which Feist argues is much more likely to have been shaped by natural selection pressures. Indeed, Feist (2001) argues that natural selection has driven mainly the more applied or technological aspects of creativity that have clear survival benefits, such as advances in science and engineering, whereas sexual selection has driven more ornamental or aesthetic aspects of creativity, including art, music, dance, and humor; forms of creativity that have come along more recently on the evolutionary scene. Feist also argues that Miller is out of touch with the creativity literature in which creativity is defined as both novel and adaptive behavior (Sternberg, 2000). Miller's focus is mostly on novel creative displays that attract the attention of potential mates. Lastly, to make the argument convincing it would be necessary to show that creative people are indeed considered more attractive, and have greater reproductive success. While there is some evidence that intelligent and creative individuals are indeed considered more attractive and have a higher number of sexual partners (Buss, 1989; Griskevicius, Cialdini, & Kenrick, 2006; Nettle & Clegg, 2006; Prokosch et al., in press), Feist notes that there is also evidence that creative people tend to be less likely to marry and when they do, have fewer children (Harrison, Moore, & Rucker, 1995), a factor that surely also impacts on reproductive success. Moreover time spent on creative projects may be time taken away from mating and child-rearing.

Additionally, Mithen (2006) presents evidence that the musicality of our ancestors and relatives did in fact have considerable survival value as a means of communicating emotions, intentions, information, and facilitating cooperation, and thus sexual selection may well not be the sole or primary selective pressure for musicality. Additionally, he notes that while it may appear at first blush that creative men have more short-term sexual partners because of their creativity, their attractiveness may be more the combination of good looks, style, and an anti-establishment persona. Mithen also points out that the finding that males produced at least ten times more music than females and were most productive around the age of 30 (in which men are in their peak mating effort and activity), could more parsimoniously be explained by the particular structure and attitudes of twentieth-century Western society.

Perhaps the most reasonable conclusion is that sexual selection helped ramp up the evolution of creativity, exaggerating certain forms, or making them not only functional but also ornamental. In this way they went beyond the realm of practicality to the realm of aesthetic functionality.

### *6.4. Non-biological Explanations for Creativity*

If culture constitutes a second form of evolution it may also exert pressures on us that differ from, or even counter, those exerted on us by our biology. The drive to create is often compared with the drive to procreate, and evolutionary forces may be at the genesis of both. In other words, we may be tinkered with by two evolutionary forces, one that prompts us to act in ways that foster the proliferation of our biological lineage, and one that prompts us to act in ways that foster the proliferation of our cultural lineage. For example, it has been suggested that we exhibit a cultural form of altruism, such that we are kind not only to those with whom we share genes but with whom we share ideas and values (Gabora, 1997). By contributing to the wellbeing of those who share our cultural makeup, we aid the proliferation of our 'cultural selves'. Similarly, when we are in the throes of creative obsession it may be that cultural forces are compelling us to give all we have to our ideas, much as biological forces compel us to provide for our children.

Note that all of the theories discussed so far in this section attempt to explain why humans are creative at all, but even with these same pressures operating we would not be particularly creative if we did not live in a richly fascinating world that *affords* creativity. Rosch (1975) provides evidence that we form concepts in such a way as to internally mirror the correlational structure of the external world. Similarly, much creativity is inspired by the goal of understanding, explaining, and/or mastering the world we live in. Thus the beauty and intricacy of our ideas, and how they unfold over time, reflects in part the beauty and intricacy of our world, not just the world we actually live in, but the potential worlds *suggested by* the world we live in, and the fact that as our internal models of the world—our worldviews—change, so does this halo of potential worlds. Indeed one could say that human creativity evolves by compelling susceptible individuals (those whose minds are poised to solve particular creatively challenging problems or engage in creative tasks) to temporarily put aside concerns associated with survival of the 'biological self', and to reach into this 'halo of possibility', rework familiar narratives, or juxtapose familiar objects and reconceptualize their inter-relationships, and thereby hone a more nuanced 'cultural self'. In sum, the creative process is compelling and our creative achievements unfold with breathtaking speed and complexity in part because we are fortunate enough to live in a world that offers infinite possibilities for exploring not just the realm of 'what is', but the realm of 'what could be'.

## 7. CONCLUSIONS

This chapter addressed a number of questions that lie at the foundation of who we are and what makes human life meaningful. Why does no other species remotely approach the degree of cultural complexity of humans? How did humans become so good at generating ideas and adapting them to new situations? Why are humans driven to create? Do creative ideas evolve in the same sense as biological life—through natural selection—or by some other means?

We began with a brief tour of the history of *Homo sapiens*, starting six million years ago when we began diverging from our ancestral apes. The earliest signs of creativity are simple stone tools, thought to be made by *Homo habilis*, just over two million years ago. Though primitive they marked a momentous breakthrough: the arrival of a species that would eventually refashion to its liking an entire planet. With the arrival of *Homo erectus* roughly 1.8 million years ago there was a dramatic enlargement in cranial capacity coinciding with solid evidence of creative thinking: task-specific stone hand axes, complex stable seasonal habitats, and signs of coordinated, long-distance hunting. It has been proposed that the larger brain allowed items encoded in memory to be more fine-grained, which facilitated the forging of creative connections between them, and paved the way for self-triggered thought, and rehearsal and refinement of skills, and thus the ability mentally go beyond 'what is' to 'what could be'.

Another rapid increase in cranial capacity occurred between 600,000 and 150,000 years ago.It preceded by some hundreds of thousands of years the sudden flourishing of creativity between 60,000 and 30,000 years ago in the Middle/Upper Paleolithic, which is associated with the beginnings of art, science, politics, religion, and probably syntactical language. The time lag suggests that behavioral modernity arose due not to new brain parts or increased memory but a more sophisticated way of *using* memory. This may have involved the onset of symbolic thinking, cognitive fluidity, and the capacity to shift between convergent and divergent or explicit and implicit modes of thought. Also, the emergence of meta-cognition enabled our ancestors to reflect on and even override their own nature.

This chapter also reviewed efforts to understand the role of creativity in not just biological but also cultural evolution. Some have investigated the intriguing possibility that the cultural evolution of ideas and inventions occurs through a Darwinian process akin to natural selection. A problem Darwinian approaches is that natural selection is inapplicable to the extent that there is inheritance of acquired traits, and so are inappropriate to the extent that individuals actively shape ideas and adapt them to their own needs and aesthetic tastes. They can account for creative change that occurs during transmission (e.g. due to biased transmission or copying error) but not change that occurs due to thinking through how something could work. Nevertheless ideas clearly exhibit phenomena observed in biological evolution such as adaptation, niches, and drift. If they do not evolve through selection, how might they evolve? It was noted that the self-organized, self-regenerating autocatalytic structures widely believed to be the earliest forms of life did not evolve through natural selection either, but through a Lamarckian process involving communal exchange of innovations. It has been proposed that what evolves through culture is individuals' internal models of the world, or worldviews, and that like early life they are self-organized and self-regenerating. They evolve not through survival of the fittest but through transformation, and they neither die nor survive intact but transform over generations as elements get incorporated and are adapted to new circumstances. Because no self-assembly code (such as the genetic code) is involved, their evolution is Lamarckian; acquired characteristics are inherited.

Finally, this chapter addressed the question of *why* creativity evolved. Some propose that creativity emerged as an evolutionary spandrel, that it promoted group bonding, or that sexual selection played an important role in shaping aesthetic/ornamental forms of creativity. Another possible answer derives from the theory that culture constitutes a second form of evolution, and that our thought and behavior are shaped by *two* distinct evolutionary forces. Just as the drive to procreate ensures that at least some of us make a dent in our biological lineage, the drive to create may enable us to make a dent in our cultural lineage. This second deeply embedded way of exerting a meaningful impact on the world and thereby feeling part of something larger than oneself may well come to be important as our planet becomes increasingly overpopulated. Thus by understanding the evolutionary origins of human creativity we gain perspective on pressing issues of today, and are in a better position to use our creativity to direct the future course of our species and our planet.

**Acknowledgements**

This work was funded in part by a grant to the first author from the *Social Sciences and Humanities Research Council of Canada*.